\documentclass[reprint,amsmath,amssymb,bm,aps,prl,superscriptaddress,floatfix]{revtex4-2}

\usepackage[pdftex]{graphicx}
\usepackage{amsmath}
\usepackage{amssymb}
\usepackage{dcolumn}
\usepackage{bm}
\usepackage{siunitx}
\usepackage[pdftex]{hyperref}
\usepackage{multirow}
\usepackage{booktabs}
\usepackage{color}
\usepackage{xcolor}
\usepackage[T1]{fontenc}
\usepackage[utf8]{inputenc}
\usepackage{float}
\usepackage{color}

\newcommand{\varParallel}{\def\@varParallel{/\kern-.2em /}%
  \mathchoice%
  {\@varParallel}%
  {\textstyle\@varParallel}%
  {\scriptscriptstyle\@varParallel}%
  {\scriptscriptstyle\@varParallel}}

\begin{document}

\preprint{APS/123-QED}

\title{Quantum oscillations in a centrosymmetric skyrmion-hosting magnet GdRu$_{2}$Si$_{2}$}

\author{N. Matsuyama}
\email{matsuyama@issp.u-tokyo.ac.jp}
\affiliation{Institute for Solid State Physics, University of Tokyo, Kashiwa, Chiba 277-8581, Japan}

\author{T. Nomura}%
\affiliation{Institute for Solid State Physics, University of Tokyo, Kashiwa, Chiba 277-8581, Japan}

\author{S. Imajo}
\affiliation{Institute for Solid State Physics, University of Tokyo, Kashiwa, Chiba 277-8581, Japan}

\author{T. Nomoto}
\affiliation{Research Center for Advanced Science and Technology, University of Tokyo, Meguro-ku, Tokyo 153-8904, Japan}
\affiliation{PRESTO, Japan Science and Technology Agency (JST), Kawaguchi, Saitama 332-0012, Japan}

\author{R. Arita}
\affiliation{Research Center for Advanced Science and Technology, University of Tokyo, Meguro-ku, Tokyo 153-8904, Japan}
\affiliation{RIKEN Center for Emergent Matter Science (CEMS), Wako, Saitama 351-0198, Japan}

\author{K. Sudo}
\affiliation{Institute for Materials Research, Tohoku University, Sendai, Miyagi 980-8577, Japan}

\author{M. Kimata}
\affiliation{Institute for Materials Research, Tohoku University, Sendai, Miyagi 980-8577, Japan}

\author{N. D. Khanh}
\affiliation{RIKEN Center for Emergent Matter Science (CEMS), Wako, Saitama 351-0198, Japan}
\affiliation{Department of Applied Physics, University of Tokyo, Bunkyo-ku, Tokyo 113-8656, Japan}

\author{R. Takagi}
\affiliation{PRESTO, Japan Science and Technology Agency (JST), Kawaguchi, Saitama 332-0012, Japan}
\affiliation{RIKEN Center for Emergent Matter Science (CEMS), Wako, Saitama 351-0198, Japan}
\affiliation{Department of Applied Physics, University of Tokyo, Bunkyo-ku, Tokyo 113-8656, Japan}
\affiliation{Institute of Engineering Innovation, University of Tokyo, Tokyo 113-0032, Japan}

\author{Y. Tokura}
\affiliation{RIKEN Center for Emergent Matter Science (CEMS), Wako, Saitama 351-0198, Japan}
\affiliation{Department of Applied Physics, University of Tokyo, Bunkyo-ku, Tokyo 113-8656, Japan}
\affiliation{Tokyo College, University of Tokyo, Bunkyo-ku, Tokyo 113-8656, Japan}

\author{S. Seki}
\affiliation{RIKEN Center for Emergent Matter Science (CEMS), Wako, Saitama 351-0198, Japan}
\affiliation{Department of Applied Physics, University of Tokyo, Bunkyo-ku, Tokyo 113-8656, Japan}
\affiliation{Institute of Engineering Innovation, University of Tokyo, Tokyo 113-0032, Japan}

\author{K. Kindo}
\affiliation{Institute for Solid State Physics, University of Tokyo, Kashiwa, Chiba 277-8581, Japan}

\author{Y. Kohama}
\affiliation{Institute for Solid State Physics, University of Tokyo, Kashiwa, Chiba 277-8581, Japan}

\date{\today}

\begin{abstract}
We have performed magnetic torque and resistivity measurements on a centrosymmetric skyrmion-host GdRu$_{2}$Si$_{2}$, in which the dominant magnetic interaction leading to skyrmion formation is under debate.
We observe both the de Haas-van Alphen and Shubnikov-de Haas oscillations in the forced ferromagnetic phase.
The angular dependence of the quantum oscillation frequencies can be reproduced by the $ab$ $initio$ calculation.
The de Haas-van Alphen oscillation is also observed in the double-$\mathbf{Q}$ phase with a different frequency to that in the forced ferromagnetic phase, indicating a Fermi surface reconstruction due to the coupling between localized spins and conduction electrons.
Based on these experimental findings, the magnetic interactions in this system are discussed.
\end{abstract}

\maketitle

\section{Introduction}\label{sec:level1}
Magnetic skyrmion has been of great interest since its discovery, because of its topological characteristics and thus the potential applications in a next-generation magnetic memory and logic device.
In the early stage of the research, skyrmions have been found in non-centrosymmetric systems, where the competition between the ferromagnetic exchange and Dzyaloshinskii-Moriya (DM) interaction favors helical spin structures \cite{roessler2006spontaneous, muhlbauer2009skyrmion, yu2010real, seki2012observation, tokunaga2015new, Kezsmarki2015, kanazawa2017noncentrosymmetric, fert2017magnetic}.
Recently, a skyrmion lattice (SkL) phase has been discovered even in several centrosymmetric rare-earth alloys that lack the DM interaction \cite{kurumaji2019skyrmion, Hirschberger2019, khanh2020nanometric, takagi2022square}.

\begin{figure}[t]
\begin{center}
\includegraphics[width=0.95\linewidth]{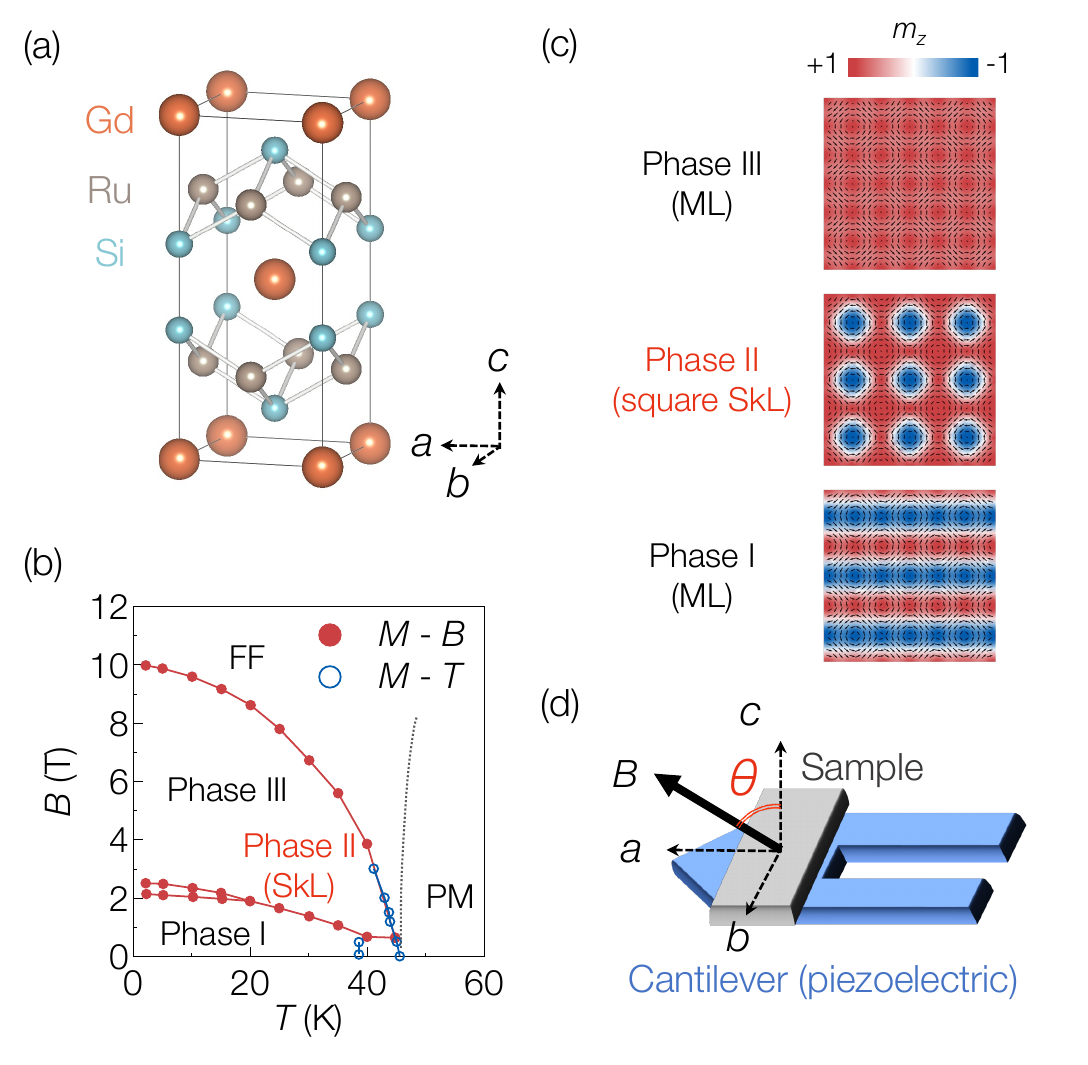}
\caption{(a) Crystal structure of GdRu$_{2}$Si$_{2}$ drawn by VESTA \cite{Momma:db5098}. (b) $B-T$ phase diagram for $B \varParallel c$ axis \cite{khanh2020nanometric}. (c) Magnetic structures in Phase I-III. (d) Schematic illustration of the experimental condition. The sample was set on the piezoelectric cantilever, and the magnetic field was rotated in the $ac$ plane. The angle $\theta$ is defined as the polar angle from the $c$ axis.}
\label{Fig1}
\end{center}
\end{figure}

GdRu$_{2}$Si$_{2}$ is one of such centrosymmetric skyrmion hosts with the shortest-period SkL ever found \cite{khanh2020nanometric}.
It crystallizes in the ThCr$_{2}$Si$_{2}$-type body-centered tetragonal structure with the space group $I4/mmm$ (Fig.~\ref{Fig1}(a)).
Gd${}^{3+}$ ions ($S=7/2, L=0$) are the source of magnetism in this material and form a square lattice within the $ab$ plane.
The local moments of Gd${}^{3+}$ show a long-range magnetic ordering at $T_{N}\sim\SI{46}{\kelvin}$ \cite{HIEBL1983287,SLASKI1984114,CZJZEK198942} with various field-induced phases \cite{GARNIER199680,samanta2008comparative}.
As shown in the phase diagram Fig.~\ref{Fig1}(b), three magnetic phases (Phase I-III) appear below $T_{N}$ with a field along the $c$ axis.
These phases are characterized by different combinations of two orthogonally-modulated spin helices with the incommensurate wavevectors $\mathbf{Q}_{1} \sim$ (0.22, 0, 0) and $\mathbf{Q}_{2} \sim$ (0, 0.22, 0) (Fig.~\ref{Fig1}(c)):
Phase I is a superposition of sinusoidal and proper-screw spin helices, and Phase III is described by two sinusoidal spin modulations, both of which are viewed as meron-antimeron lattice states with different patterns \cite{DA, khanh2020nanometric,khanh2022zoology}.
Phase II is the square SkL state with two proper-screw spin structures.
Above $\SI{10}{\tesla}$, in the forced ferromagnetic (FF) phase, the magnetization saturates at 7$\mu_{\mathrm{B}}$ per Gd${}^{3+}$.

The skyrmion stabilization mechanism for this material is still under debate.
In the centrosymmetric crystal structure, the conventional mechanism based on the DM interaction is forbidden.
Incommensurate magnetic modulations in rare-earth intermetallics are often brought about by the indirect exchange interaction mediated by conduction electrons, the so-called Ruderman-Kittel-Kasuya-Yosida (RKKY) interaction \cite{jensen1991rare}.
Indeed, the coupling between conduction electrons and localized magnetic moments has been experimentally confirmed in GdRu$_{2}$Si$_{2}$ by the real-space observation using the scanning-tunneling microscope (STM) technique \cite{yasui2020imaging}.
Theoretically, several spin models based on the anisotropic exchange interactions \cite{PhysRevB.103.104408,PhysRevB.103.064414}, the RKKY interaction \cite{PhysRevLett.124.207201,Yambe2021,PhysRevB.104.184432}, and the multiple-spin interactions mediated by itinerant electrons \cite{PhysRevLett.101.156402,PhysRevLett.108.096401,PhysRevLett.118.147205,hayami2017effective,PhysRevB.103.024439,PhysRevB.103.054422}, have succeeded in reproducing SkL state in centrosymmetric systems.
As for the microscopic origin to stabilize the fundamental helical modulation, there are two different interpretations proposed by first-principles calculations.
One suggests the RKKY interaction enhanced by the Fermi-surface nesting stabilizes the zero-field helical modulation \cite{PhysRevLett.128.157206}, while the other proposes the competition between ferromagnetic interaction in Gd-5$d$ channel and antiferromagnetic interaction in Gd-4$f$ channel, namely, the interorbital frustration plays a crucial role \cite{PhysRevLett.125.117204}.
In order to discuss the possible origin of the helical modulation in this centrosymmetric $f$-electron system, the electronic band structure of GdRu$_{2}$Si$_{2}$ needs to be clarified, which reflects how the $f$ and conduction electrons interact with each other and contribute to the magnetic structures.

Here, we perform magnetic torque and resistivity measurements under high magnetic fields and at low temperatures to reveal the electronic structure of GdRu$_{2}$Si$_{2}$ through the de-Haas van-Alphen (dHvA) and Shubnikov de-Haas (SdH) oscillations.
The angular dependence of the dHvA-oscillations frequency in the FF phase agrees with the $ab$ $initio$ calculation.
Furthermore, the Fermi surface reconstruction from Phase III to the FF phase is observed, which indicates the coupling between conduction electrons and localized spins.
While the plausibility of the interorbital frustration mechanism should be separately investigated, these experimental findings suggest that the RKKY interaction is relevant to understand the helical magnetism and skyrmion formation in this system.

\section{Methods}
High-quality single crystals were grown by the floating zone method under an Ar gas flow \cite{khanh2020nanometric}.
Several pieces of the sample with residual resistivity ratio (RRR) in the range of 50 to 100 were used in this research.

The angular dependence of the magnetic torque ($\tau$) was investigated with a two-axis rotator up to $\SI{18}{\tesla}$ using a superconducting magnet in IMR, Tohoku Univ. and with a single-axis plastic rotator up to $\SI{58}{\tesla}$ using a non-destructive pulsed magnet in ISSP, Univ. of Tokyo.
For both measurements, we utilized a commercial piezoelectric cantilever \cite{ohmichi2002torque, PhysRevB.101.224509}.
A tiny single crystal with the size of $\sim 200 \times 100 \times \SI{30}{\micro\metre^3}$ was mounted on the top of the cantilever (Fig.~\ref{Fig1}(d)).
Magnetic torque ($\tau$), defined as the cross product of magnetization and magnetic fields ($\bm{\tau} = \bm{M} \times \bm{H}$), were measured as the change in the resistance of the piezoelectric cantilever.

Resistivity ($\rho$) of GdRu$_{2}$Si$_{2}$ was measured by the four-probe method with the current along the $ab$ plane.
The angular dependence of $\rho$ was also measured with the two-axis rotator up to $\SI{18}{\tesla}$ at IMR, Tohoku Univ.
The field direction was rotated in the $ac$ plane, and the field angle $\theta$ was defined as the relative angle from the $c$ axis (Fig.~\ref{Fig1}(d)).


\begin{figure}[t]
\centering
\includegraphics[width=0.95\linewidth]{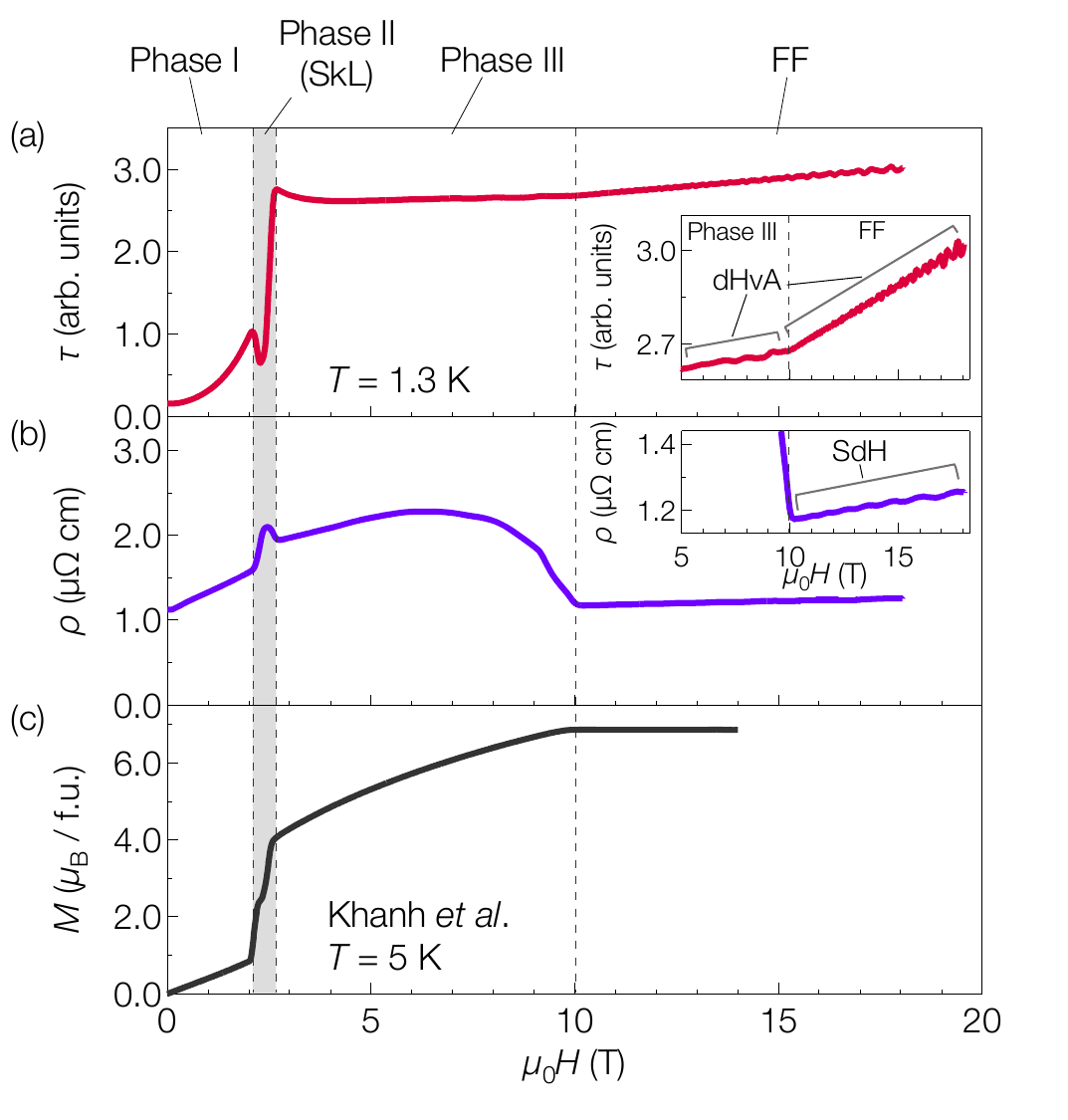}
\caption{Field dependence of (a) magnetic torque ($\theta = \SI{2}{\degree}$), (b) resistivity ($B \varParallel c$) at $T=\SI{1.3}{\kelvin}$, and (c) magnetization ($B \varParallel c$) at $T=\SI{5}{\kelvin}$ \cite{khanh2020nanometric}, of GdRu$_{2}$Si$_{2}$. Observed quantum oscillations are highlighted in the insets.}
\label{Fig2}
\end{figure}

We compared the experimental data with the calculated electronic band structures.
The calculation details are as follows:
First, we performed spin density functional theory calculations based on the projector augmented wave scheme implemented in the Vienna \textit{ab initio} simulation package \cite{PhysRevB.54.11169,PhysRevB.59.1758}.
Here, we employed the Perdew-Burke-Ernzerhof exchange-correlation functional
\cite{PhysRevLett.77.3865} and assumed the collinear ferromagnetic order.
The strong interactions in 4$f$ orbitals of Gd were taken into account by the DFT+$U$ method, where we set  $U= \SI{6.7}{\electronvolt}$ and $J= \SI{0.7}{\electronvolt}$.
Then, we constructed the tight-binding model based on the Wannier function via the WANNIER90 code \cite{MOSTOFI2008685,Pizzi_2020} and interpolated energy eigenvalues on the dense $k$-grid of $51 \times 51 \times 51$.
Finally, we calculated the dHvA frequencies using the SKEAF code \cite{julian2012numerical}.

\begin{figure*}[t]
\centering
\includegraphics[width=\linewidth]{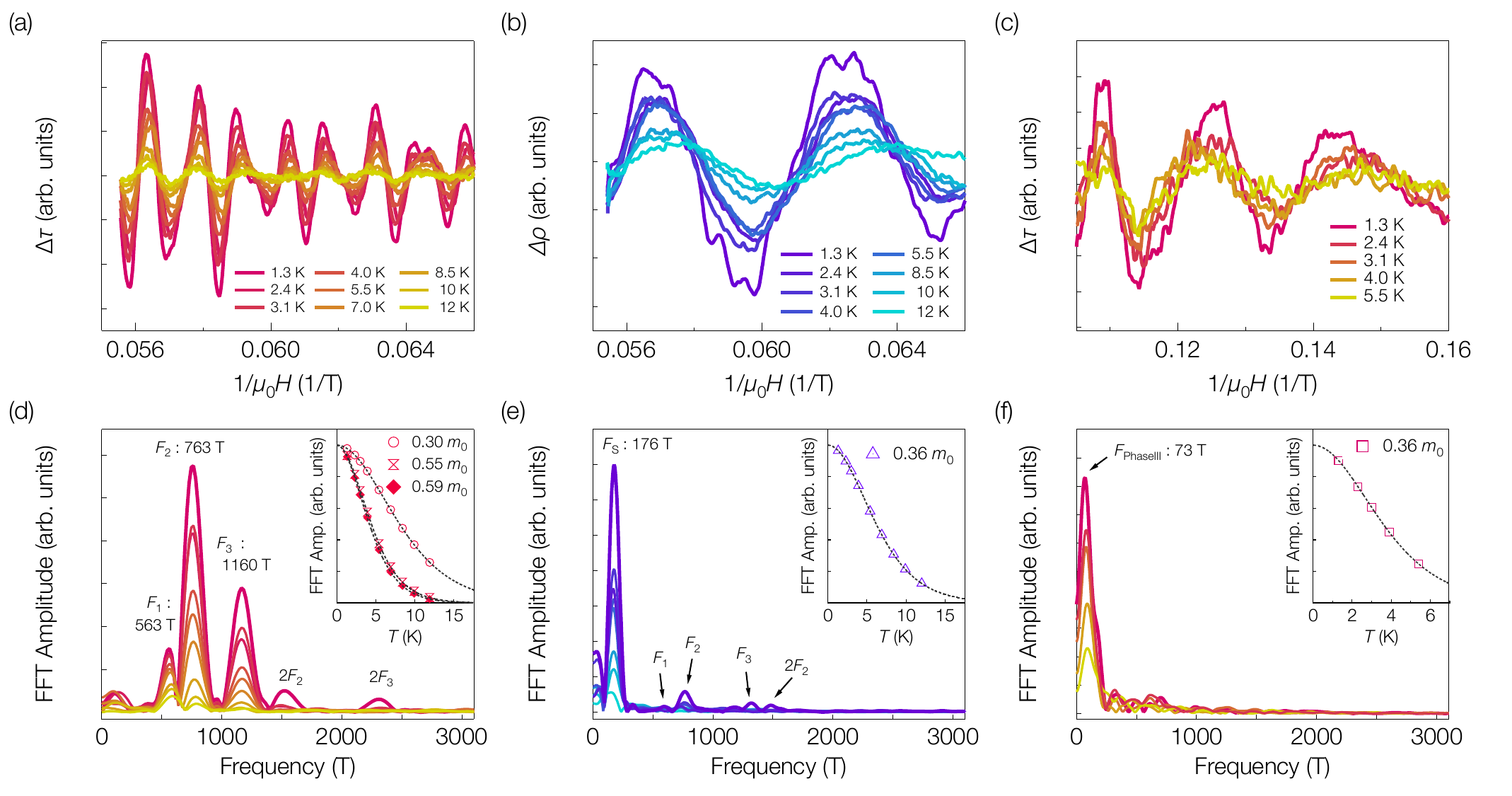}
\caption{Temperature dependence of oscillating components (top) and FFT spectra (bottom) of the observed dHvA(a, d) and SdH(b, e) in the FF phase and dHvA in Phase III(c, f). The field angle for dHvA and SdH experiments are $\theta = \SI{2}{\degree}$ and $\theta = \SI{0}{\degree}$ ($B \varParallel c$), respectively.}
\label{Fig3}
\end{figure*}

\begin{table*}
\caption{Oscillation frequency $F$ (T), Fermi wavevector $k_{F}$ ({\AA}${}^{-1}$), effective mass $m^{*}/m_{0}$, Dingle temperature $T_D$ (K), scattering time $\tau_{q}$ (ps), and mean free path $l_{q}$ (nm) for each dHvA and SdH oscillation observed in the experiment in GdRu$_{2}$Si$_{2}$ ($B \varParallel c$).}
\label{tab0}
\begin{ruledtabular}
\begin{tabular}{cccccccc}
                                 &                           & $F$ (T) & $k_{F}$ ({\AA}${}^{-1}$) & $m^{*}/m_{0}$ & $T_D$ (K)  & $\tau _ {q}$(ps) & $l_{q}$ (nm) \\ \hline
\multirow{3}{*}{dHvA (FF phase)} & $F_{1}$                   & 563     &  0.13                 & 0.30          & 15         &  0.081           & 40           \\
                                 & $F_{2}$                   & 763     &  0.15                 & 0.55          & 6.8        &  0.18            & 57           \\
                                 & $F_{3}$                   & 1160    &  0.19                 & 0.59          & 7.5        &  0.16            & 59           \\ \hline
SdH (FF phase)                   & $F_{\mathrm{S}}$          & 176     &  0.073                & 0.36          & 9.2        &  0.13            & 30           \\ \hline
dHvA (Phase III)                 & $F_{\mathrm{Phase III}}$  &  73     &  0.047                & 0.36          & 1.3        &  0.23            & 35\\
  \end{tabular}
\end{ruledtabular}
\end{table*}

\section{Results}
\subsection{A. Magnetic torque and magnetoresistance for $B \varParallel c$}

Figures~\ref{Fig2}(a) and \ref{Fig2}(b) show the field dependences of $\tau$ with the field \SI{2}{\degree} away from the $c$ axis and $\rho$ for $B \varParallel c$.
These data are obtained at $\SI{1.3}{\kelvin}$.
Magnetization $M$ at $\SI{5}{\kelvin}$ reported in Ref.~\cite{khanh2020nanometric} is shown for comparison (Fig.~\ref{Fig2}(c)).
All three quantities show several anomalies under magnetic fields corresponding to magnetic phase transitions.
Below $\SI{2.0}{\tesla}$ (Phase I), magnetization and resistivity increase almost linearly, while magnetic torque shows a nearly quadratic change ($\bm{\tau} = \bm{M} \times \bm{H} \propto \bm{H}^2$).
The magnetization shows a step-like structure in Phase II (SkL phase; $2.0 < B < \SI{2.7}{\tesla}$), which is observed as a dip in magnetic torque and a hump in resistivity.
The hump structures in the longitudinal and Hall resistivity of GdRu$_{2}$Si$_{2}$ have been reported within this field range \cite{khanh2020nanometric}.
The dip in magnetic torque suggests a drastic change of magnetic anisotropy in Phase II related to the SkL formation.
In Phase III, magnetization exhibits a sublinear increase and saturates at $\SI{10}{\tesla}$.
With increasing the magnetic field, magnetic torque shows a nonlinear increase up to the saturation, which may reflect the sublinear increase of the magnetization.
The resistivity shows a broad peak and decreases towards saturation.
The decrease in resistivity can be attributed to the suppression of spin fluctuations at high magnetic fields \cite{samanta2008comparative}.

Quantum oscillations are observed both in magnetic torque (dHvA) and resistivity (SdH) in the FF phase ($B>\SI{10}{\tesla}$) as shown in the insets of Fig.~\ref{Fig2}(a) and (b), respectively.
Figures~\ref{Fig3}(a) and (b) show the oscillatory components of magnetic torque and magnetoresistance at various temperatures as a function of inverse fields.
The oscillatory components of the observed quantum oscillations are obtained by subtracting the smooth background deduced from a cubic polynomial fit.
The fast Fourier transform (FFT) spectra of the dHvA and SdH oscillations are shown in Figs.~\ref{Fig3}(d) and (e), respectively.
The FFT spectrum of the dHvA oscillations in the FF phase shows three main peaks $F_{1}=\SI{563}{\tesla}$, $F_{2}=\SI{763}{\tesla}$, $F_{3}=\SI{1160}{\tesla}$ and additional small peaks at higher frequency region, some of which are harmonics of the main peaks (Fig.~\ref{Fig3}(d)).
In contrast to the dHvA oscillations, the SdH oscillation consists of one predominant frequency  $F_{\textrm{S}}=\SI{176}{\tesla}$ (Fig.~\ref{Fig3}(e)).
The peaks observed in the dHvA oscillations are seen as tiny peaks indicated by arrows.
According to the Onsager relation $F = (\hbar/2\pi e)S_{F}$, where $S_{F}$ is a cross-sectional area in momentum-space, assuming a circular cross-section $S_{F} = \pi k_{F}^2$ ($k_{F}$ : Fermi wavevector), one can estimate $k_{F}$ corresponding to the observed frequency.
The quantum oscillation frequencies and estimated Fermi wavevectors for $B \varParallel c$ are listed in Table~\ref{tab0}.

We analyze the quantum oscillations of magnetic torque and resistivity using the Lifshitz-Kosevich (LK) formula expressed as
\begin{eqnarray}
  \Delta\tau &=&  C \sum_{p=1}^{\infty} B^{3/ 2}  R_{T} R_{D} \sin \left[2 \pi \left(\frac{F}{B}-\frac{1}{2}\right)+\Phi\right],\nonumber\\
  \Delta\rho/\rho_{0} &=& C \sum_{p=1}^{\infty} B^{1 / 2}  R_{T} R_{D} \cos \left[2 \pi \left(\frac{F}{B}-\frac{1}{2}\right)+\Phi\right],\nonumber
\end{eqnarray}
where $C$ is a constant \cite{MOIM}.
The Landau-level broadening caused by temperature and impurity scattering is represented by the damping factors $R_{T}=(K\mu /B) / [\sinh(K\mu T/B)]$ and $R_{D}=\mathrm{exp}(-K\mu T_{D}/B)$, respectively, with $K=(2\pi^2 m_0 k_B)/(e \hbar)$ and $\mu=m^{*}/m_{0}$, where $k_{B}$, $m^{*}$ and $T_{D}$ represent the Boltzmann-constant, the effective cyclotron mass, and Dingle temperature.
$T_{D}$ relates to the impurity scattering rate and corresponding scattering time $\tau_{q}$ can be deduced by $\tau_{q}=\hbar  / (2 \pi k_{B} T_{D})$.
The phase factor $\Phi$ derives from the dimensionality of the Fermi surface, Zeeman splitting, and Berry phase \cite{PhysRevLett.82.2147}.
The observation of the quantum limits is often necessary for the precise determination of $\Phi$, which is not achieved in this research.
By fitting $R_{T}$ and $R_{D}$ to the temperature (insets of Figs.~\ref{Fig3}(d) and (e)) and field dependences of the FFT amplitude, we calculated $m^{*}$, $T_D$, $\tau_{q}$,
and mean-free path $l_{q} = v_{\mathrm{F}} \cdot \tau_{q} = \hbar^2 k_{\mathrm{F}} / (2\pi k_{B} m^{*} T_{D})$ ($v_{\mathrm{F}} = \hbar k_{\mathrm{F}} / m^{*}$ : Fermi velocity) as summarized in Table~\ref{tab0}.

The dHvA oscillation is observed in the high-field region of Phase III ($6.0<B<\SI{10}{\tesla}$), as well.
The oscillatory components and the FFT spectra of them are shown in Fig.~\ref{Fig3}(c) and Fig.~\ref{Fig3}(f), respectively.
The main peak locates at $\SI{73}{\tesla}$, which is roughly one order of magnitude smaller than the dHvA frequencies observed in the FF phase.
The fit to the LK formula yields $m^* \sim 0.36 m_{0}$ (inset of Fig.~\ref{Fig3}(f)), and $T_{D} \sim \SI{5.3}{\kelvin}$.
$\tau_{q}$ and $l_{q}$ are estimated as $\SI{0.23}{\pico\second}$ and $\SI{35}{\nano\metre}$.

\begin{figure}[t]
\centering
\includegraphics[width=\linewidth]{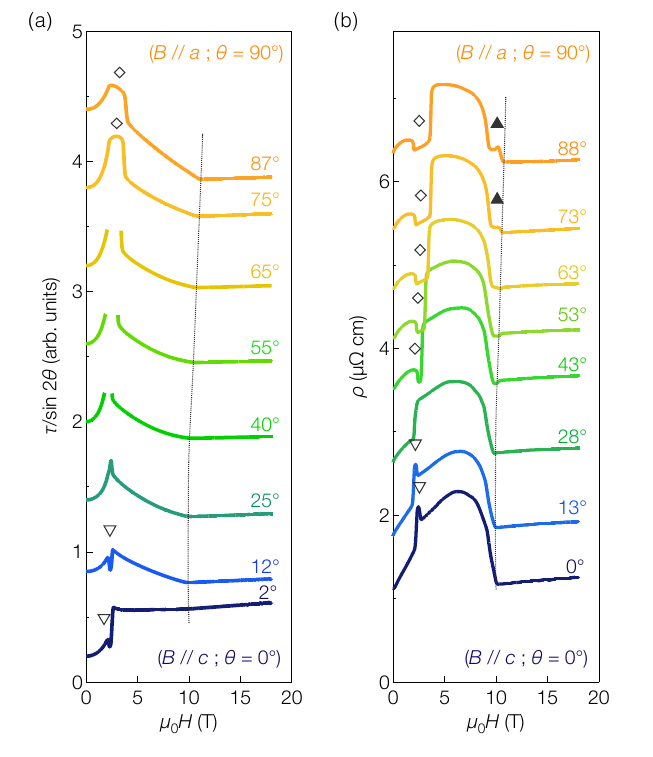}
\caption{Angular dependences of (a) magnetic torque and (b) magnetoresistance at $T=\SI{1.3}{\kelvin}$. The data are shown only for selected angles with constant shifts for clarity.}
\label{Ang}
\end{figure}

\subsection{B. Angular dependence}
\begin{figure*}[t]
\centering
\includegraphics[width=1.03\linewidth]{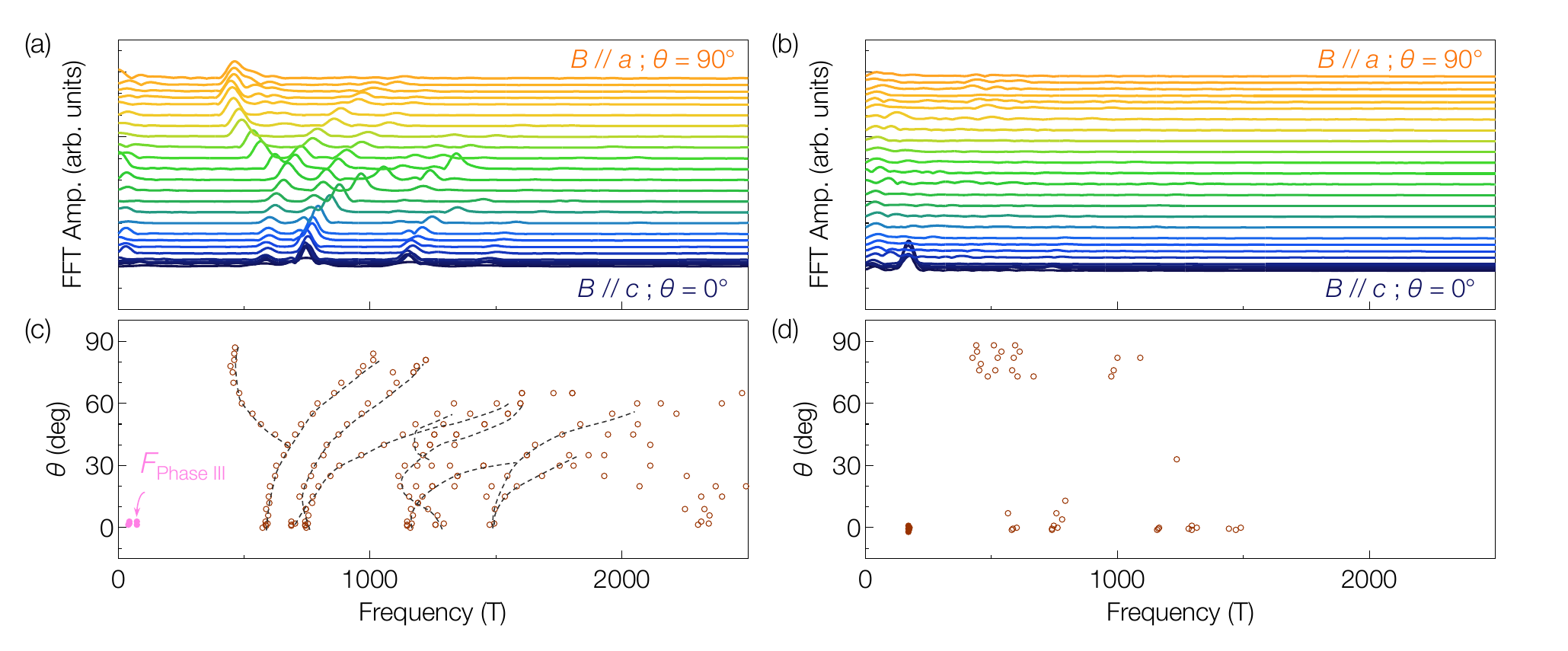}
\caption{Angular dependence of the (a)dHvA and (b)SdH oscillations in the FF phase. The peak positions of the spectra are plotted in (c) and (d), respectively.}
\label{WF}
\end{figure*}
Next, the angular dependences of the magnetoresistance and magnetic torque are addressed.
Figure~\ref{Ang}(a, b) shows the angular dependences of the magnetic torque and magnetoresistance at $T = \SI{1.3}{\kelvin}$.
The torque scaled by $\sin{2 \theta}$ is shown to magnify the torque signal around $B \varParallel a$ and $c$, and the saturation fields are highlighted with dotted lines.
At some angle $\theta$ and field ranges, we omit the torque data which are out of the measurement range.
The dip in $\tau$ and hump in $\rho$, which correspond to Phase II, are observed around $\theta = \SI{0}{\degree}$ (open inverted triangle).
They vanish when the field direction is tilted larger than $\theta \sim \SI{25}{\degree}$.
The application of the tilted magnetic field with $\theta > \SI{40}{\degree}$ results in the appearance of another step-like feature in both $\tau$ and $\rho$ (open diamond).
These features suggest the disappearance of the SkL phase and stabilization of a different magnetic phase for the tilted field directions.
When the field is applied in the vicinity of the $a$ axis ($\theta > \SI{70}{\degree}$), an additional hump structure appears only in magnetoresistance (filled triangle), which also indicates an appearance of another magnetic phase just below the saturation field.
The high-field phase is not detected in the torque measurements, which may infer similar magnetic anisotropy for these adjacent phases.
The angular dependence of the $B$-$T$ phase diagram has been reported in Ref.~\cite{khanh2022zoology}, which is consistent with our observations.

The dHvA oscillations are clearly observed in the FF phase regardless of the field direction, while the dHvA in Phase III and SdH in the FF phase are observed in the limited angle range.
Figures~\ref{WF}(a) and \ref{WF}(b) show the angular dependences of the normalized FFT spectra of dHvA and SdH oscillations in the FF phase, respectively.
On each bottom, peak positions in the FFT spectra are plotted as a function of the field angle $\theta$ (Figs.~\ref{WF}(c) and \ref{WF}(d)).
The dotted curves are drawn in Fig.~\ref{WF}(c) for the eye guide, where several branches can be identified.
The $F_{\textrm{S}}$ peaks observed in SdH oscillations (shown as filled symbols in Fig.~\ref{WF}(d)) are seen only in the $\SI{3}{\degree}$ range from the $c$ axis.
This might be because of the quasi-two-dimensional crystal structure of this sample: Gd magnetic layers are sandwiched between two (Ru, Si) blocks.
When the field is tilted away from the $c$ axis, the scattering probability in the cyclotron orbit may be enhanced with a stacking fault.

The dHvA oscillation in Phase III also appears only in the narrow-angle range $|\theta| < \SI{3}{\degree}$ as shown in pink dots in Fig.~\ref{WF}(c).
According to the previous study \cite{khanh2022zoology}, the phase boundary does not change by $\SI{3}{\degree}$ tilting, indicating that the disappearance of the dHvA is not related to any magnetic phase transition.
Alternatively, we propose that the weak out-of-plane spin correlation due to the quasi-two-dimensional structure accounts for the disappearance of the quantum oscillation in the tilted field direction.
The weak spin correlation between the layers may result in an irregular phase change of the spin modulation across the layers.
This increases the scattering probability for interlayer conduction and could lead to a sudden reduction of the quantum oscillation magnitude when the field direction is tilted.
Indeed, the weak spin coupling is suggested by the theoretical study \cite{PhysRevLett.125.117204}.


\subsection{C. Torque measurement \\under the pulsed magnetic fields}
\begin{figure}
\centering
\includegraphics[width=1.05\linewidth]{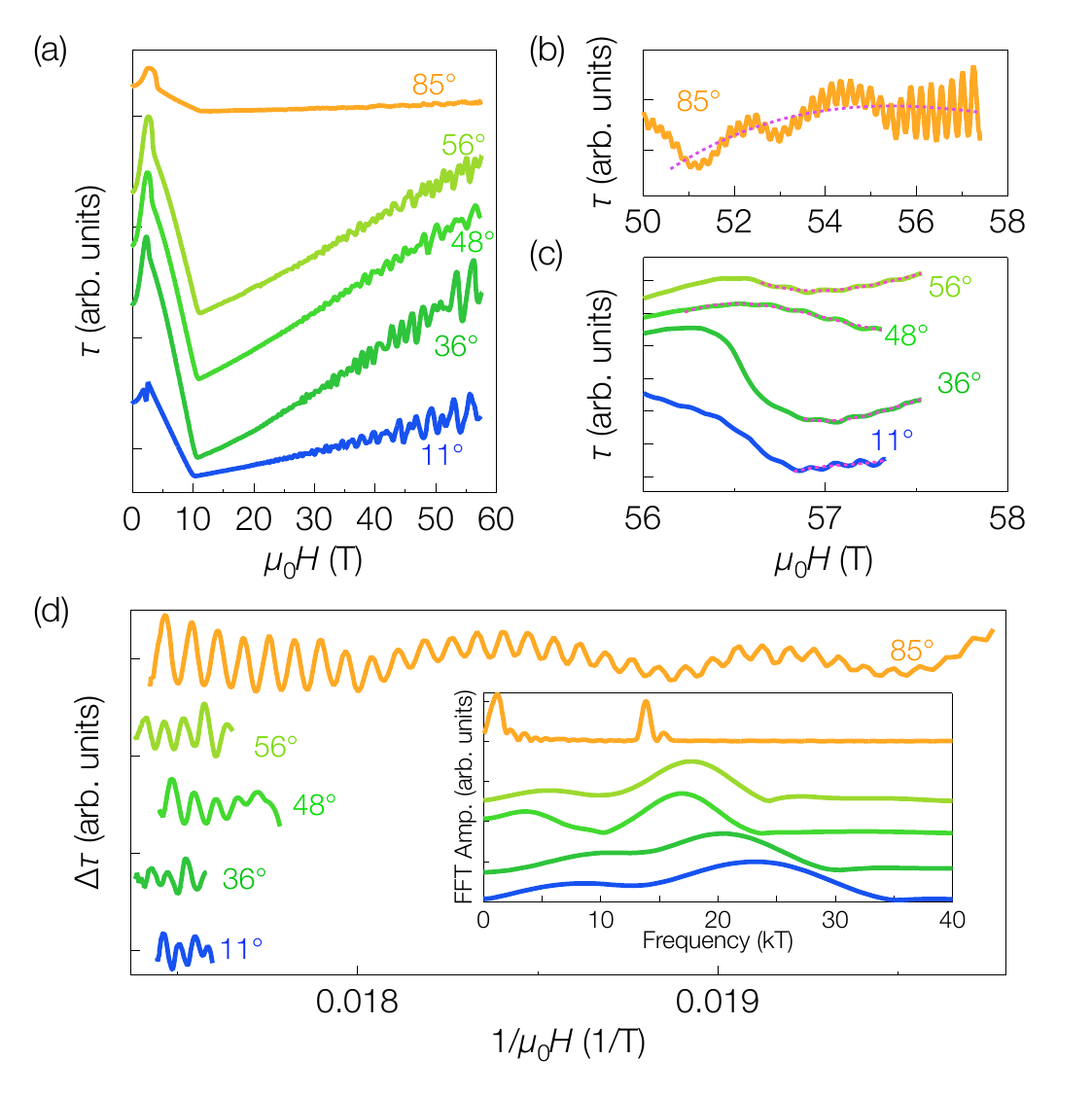}
\caption{(a) Field dependence of magnetic torque of GdRu$_{2}$Si$_{2}$ using the pulsed field up to 58 T with constant shifts. (b,c) Expanded view of the high-field region of (a). Subtracted backgrounds for each data are shown in pink dotted lines. (d) The oscillatory components of the high-frequent dHvA and the FFT spectra of the dHvA oscillations (inset), both shifted for clarity.}
\label{pulse}
\end{figure}

The field dependences of magnetic torque measured in pulsed fields up to $\SI{58}{\tesla}$ are shown in Fig.~\ref{pulse}(a).
At $\sim \SI{56}{\tesla}$, another high-frequency oscillation starts (Fig.~\ref{pulse}(c)), which is not observed in the data up to $\SI{18}{\tesla}$.
For the data taken at $\theta = \SI{85}{\degree}$ in Fig.~\ref{pulse}(a), the high-frequency oscillation is observed above $\sim \SI{50}{\tesla}$ with the lower-frequency one observed in the steady-field experiment.
Here, we focus on the high-frequency oscillation which is only observed in the pulsed-field experiments (Fig.~\ref{pulse}(b)).
To extract the high-frequency oscillations, we subtract the smooth background obtained by a cubic polynomial fit (dashed pink curve in Figs.~\ref{pulse}(b) and \ref{pulse}(c)) from each of the high-field $\tau$ data.
The oscillatory components of the high-frequency oscillation and FFT spectrum are shown in Fig.~\ref{pulse}(d) and the inset of Fig.~\ref{pulse}(d).
Because $\tau (B)$ for $\theta = \SI{85}{\degree}$ contains a large number of oscillations, a sharp double-peak structure centered at $\sim \SI{15000}{\tesla}$ is resolved as seen in the inset of Fig.~\ref{pulse}(d).
The origin of this fine structure is considered to be the spin-split bands as discussed later.
As seen in the inset of Fig.~\ref{pulse}(d), the low-frequency oscillation also induces a broad peak structure at about $\SI{1200}{\tesla}$, which is not well resolved due to the limited field window used for the FFT.
In contrast to the data for $\theta = \SI{85}{\degree}$, $\tau (B)$ for the other angles shows a too-small number of oscillations to resolve the fine structure of the FFT peak.

\subsection{D. Theoretical calculation}
\begin{figure}
\centering
\includegraphics[width=0.98\linewidth]{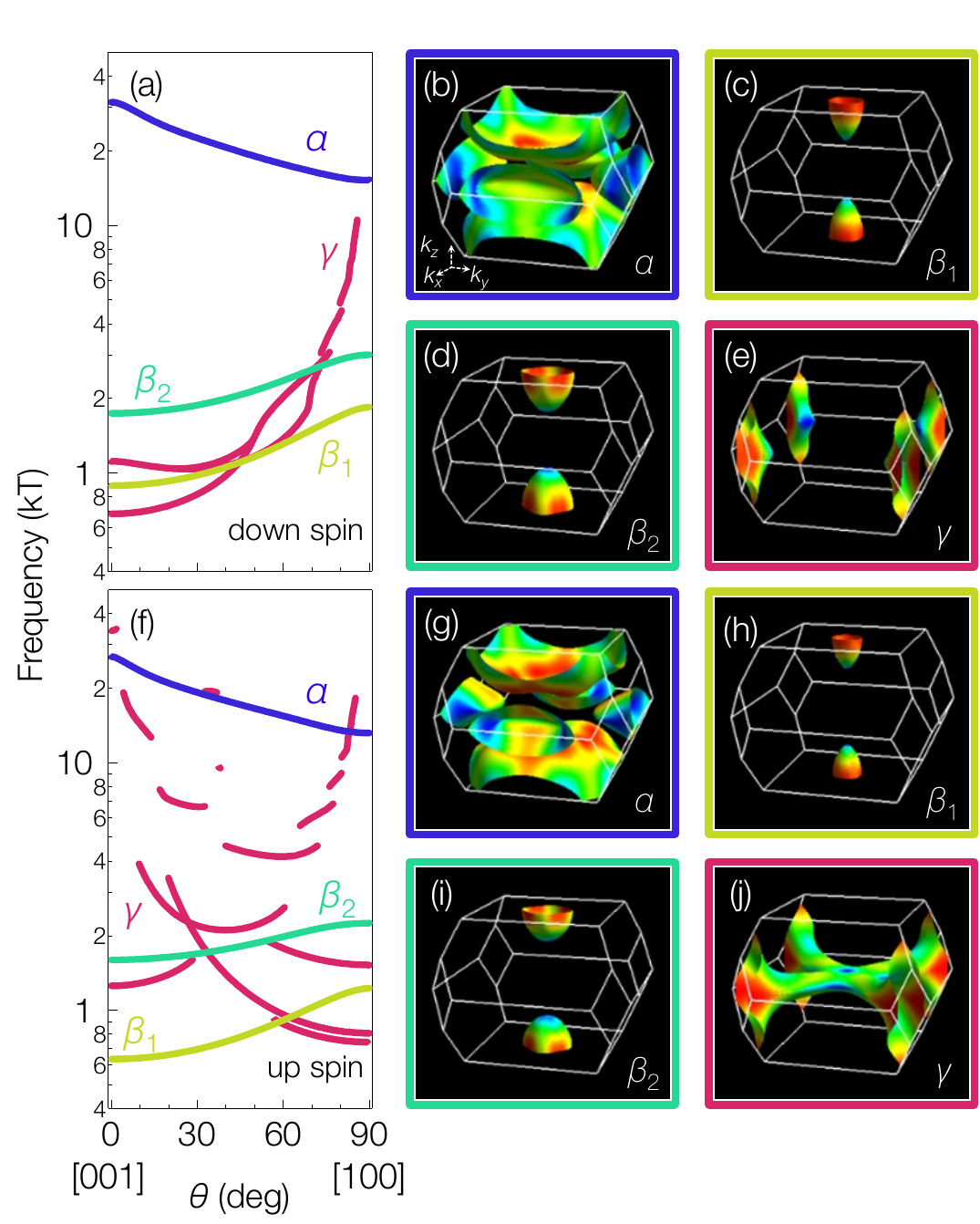}
\caption{(a,f) Calculated frequencies of the dHvA oscillations for GdRu$_{2}$Si$_{2}$ as a function of the magnetic field direction. (b-e,g-j) Fermi surfaces for down spin (b-e) and up spin (g-j) drawn by the FermiSurfer code \cite{KAWAMURA2019197}. $\alpha$ and $\beta_\mathrm{1,2}$ are hole sheets, while $\gamma$ is an electron sheet.}
\label{FigTheory}
\end{figure}

The calculated Fermi surfaces and the angular dependences of the quantum oscillations are shown in Fig.~\ref{FigTheory}.
Four Fermi surfaces are expected for each spin-up and spin-down band.
There are one large hole surface termed $\alpha$ and two small hole pockets termed $\beta_\mathrm{1,2}$ centered at the \textit{Z} point.
In addition, one electron pocket termed $\gamma$ is predicted.
Note that the topology of $\gamma$ surfaces depends on its spin states because of the large spin polarization.
The density of states around the zone center exist only for the spin-up band (Fig.~\ref{FigTheory}(j)).
In contrast, the pillar-like parts at the corners are split into two Fermi surfaces for the spin-up and spin-down bands (Fig.~\ref{FigTheory}(e) and (j)).

The calculation predicts that the topology of the $\gamma$ surface of the up spins changes when the Fermi energy is shifted by $\SI{50}{\milli\electronvolt}$.
A hole emerges at the zone center, where the Fermi surface changes from an ellipsoidal sphere to a torus by shifting the Fermi energy.
Similar behavior is also predicted for the non-$f$ reference material LaRu$_{2}$Si$_{2}$ \cite{suzuki2010change,PhysRevB.65.035114}.

\section{Discussions}
\subsection{A. Fermi surfaces of GdRu$_{2}$Si$_{2}$}
Here, to clarify the electronic structure in the FF phase, we compare the angular dependence of the calculated oscillation frequencies with that of dHvA oscillations, which are observed throughout the angle range.

\begin{figure}
\centering
\includegraphics[width=0.9\linewidth]{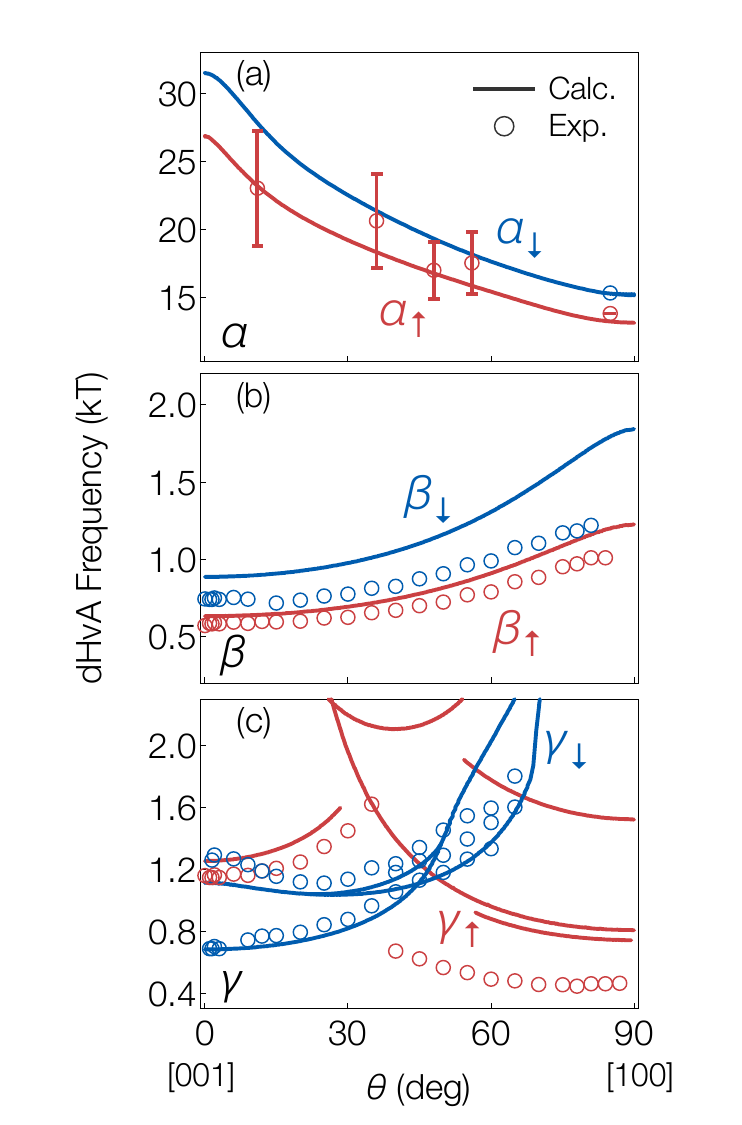}
\caption{Angular dependence of the dHvA-oscillations frequencies for (a) $\alpha$ band, (b) $\beta$ band, and (c) $\gamma$ band. }
\label{FigAngCompare}
\end{figure}

We identify six branches; one electron sheet and two hole sheets (three for each spin-split subband).
In Fig.~\ref{FigAngCompare}, the calculated FFT frequencies (solid lines) are plotted with the experimental data (circles).
First, the calculated angle dependence of the large hole surface ($\alpha$) shows a good agreement with the angular dependence of the high-frequency oscillation detected in the pulsed magnetic fields.
The splitting width of the double-peak structure resolved for $\theta = \SI{85}{\degree}$ is consistent with the frequency difference in the calculated spin-up and -down branches (Fig.~\ref{FigAngCompare}(a)), although the large error for $\theta < \SI{85}{\degree}$ hinders to track the angular dependence of the spin splitting.
Next, we find that the angular dependences of $F_{1}$ and $F_{2}$ resemble the calculated angular dependences for the spin-split ellipsoidal hole surface $\beta$ as seen in Fig.~\ref{FigAngCompare}(b).
The slight discrepancy between the experimental and theoretical results is the number of Fermi surfaces.
The present calculation predicts four fundamental oscillations originating from the two spin-split pairs of ellipsoidal hole surfaces $\beta_\mathrm{1,2}$, while the torque magnetometry detects only two dHvA oscillations originating from one spin-split pair of $\beta_\mathrm{1}$.
The difference in the number of the ellipsoidal hole surface $\beta$ was also discussed in several calculation studies for LaRu$_{2}$Si$_{2}$, CeRu$_{2}$Si$_{2}$ \cite{suzuki2010change,PhysRevB.65.035114,PhysRevB.51.10375}, and another
ThCr$_{2}$Si$_{2}$-type material \cite{reiss2013lurh2si2}.
According to these studies, the number of the sheets $\beta$ are sensitive to the Fermi energy and the Wyckoff parameter of Si atoms.
Thus, a detailed investigation of the lattice parameters at low temperatures, $e.g$., by diffraction experiments, might help to improve the accuracy of the ellipsoidal sheets $\beta$.
Lastly, the FFT frequencies including  $F_{3}$ are found to be consistent with the angular dependence of the calculated angular dependence for the electron sheets $\gamma$ as shown in Fig.~\ref{FigAngCompare}(c).
In summary, although there is a slight difference in the number of hole sheets $\beta$, experimental data are qualitatively reproduced by the calculated band structures.

Next, we discuss the SdH oscillation observed with $F_{\mathrm{S}} = \SI{176}{\tesla}$, which is not observed in the dHvA oscillations.
Such a different sensitivity of thermodynamic and transport probes or of different transport probes on the quantum oscillation has been reported in earlier works \cite{PhysRevLett.105.216409, PhysRevLett.117.206401}.
The SdH oscillation at $F_{\mathrm{S}} = \SI{176}{\tesla}$ might be related to the $\gamma$ surface of the up spins ($\gamma_{\uparrow}$) (Fig.~\ref{FigTheory}(j)).
The electronic structure of this band varies with a slight shift in the Fermi level, and it can be either an ellipsoidal sphere or a torus.
When this Fermi surface becomes a torus, the quantum oscillation corresponding to the inner core of the torus can appear, which is a candidate for the $F_{\mathrm{S}}$ branch.
In contrast to the transport probes sensitive to both the variation of the density of states and the change in the scattering time, the thermodynamic probes are only sensitive to the variation of the density of states \cite{PANTSULAYA1989317}.
Thus, the small cross-section of the inner core might have only a small effect on the density of states and show negligibly small dHvA oscillation, while it can affect the scattering time which results in a larger SdH oscillation.

Another possible explanation is the quantum interference effect, which is not due to Landau quantization but results from electrons on different trajectories \cite{stark1974interfering, stark1977comparison}.
This effect causes an oscillation only in resistivity and not in other thermodynamic properties, as is the present case.
The amplitude of this oscillation, however, is known to be insensitive to variations in temperature \cite{PhysRevLett.26.556,PhysRevResearch.2.023217}.
This is in contrast to the observed oscillation of $F_{\mathrm{S}}$ branch.
A similar difference between dHvA and SdH is also reported for CeAgSb${}_2$ \cite{PhysRevB.72.104428}.
In this quasi-two-dimensional 4$f$-localized ferromagnet, the electronic structure has been investigated by both dHvA \cite{doi:10.1080/13642810208222948} and SdH effects \cite{PhysRevB.72.104428,PhysRevB.75.014413,PhysRevB.60.13371}.
According to the SdH investigation, the main peak was found at $\SI{25}{\tesla}$, which had not been reported in the dHvA experiment nor predicted by the first-principles calculation.
Furthermore, this SdH oscillation is sensitive to the field direction and strongly suppressed when the field is tilted more than $\SI{7}{\degree}$ away from the $c$ axis, while the dHvA oscillations are observed regardless of the field direction.
These features are discussed in terms of field-orientation-sensitive scattering due to the magnetic anisotropy, the dimensionality of the Fermi surface, and quantum interference \cite{PhysRevB.72.104428}.

\subsection{B. Comparison with LaRu$_{2}$Si$_{2}$}
Here, the observed Fermi surfaces in GdRu$_{2}$Si$_{2}$ are compared with those of the non-$f$ reference material LaRu$_{2}$Si$_{2}$ \cite{doi:10.1143/JPSJ.61.960,doi:10.1143/JPSJ.61.2388,suzuki2010change}.
Table~\ref{tab} and \ref{tab2} compares the observed dHvA frequencies and corresponding cyclotron masses in the two materials for $B \varParallel c$ and $B \varParallel a$, respectively.
Note that the temperature dependence of dHvA was investigated with the field angle $\theta = \SI{2}{\degree}$ (Figs.~\ref{Fig3}(a,d)), which is adopted to Table~\ref{tab}.
In LaRu$_{2}$Si$_{2}$, a large hole sheet, two small hole ellipsoidal sheets, and one electron sheet are experimentally reported.
The overall electronic structures of both compounds are almost the same and consist of $\alpha$, $\beta$, and $\gamma$, although the number of hole ellipsoidal sheet $\beta$ is different between GdRu$_{2}$Si$_{2}$ (one each for spin-up and -down) and LaRu$_{2}$Si$_{2}$ (two spin-degenerated bands).
The major difference of GdRu$_{2}$Si$_{2}$ from LaRu$_{2}$Si$_{2}$ is the spin polarization in the $\gamma$ band in the FF state.
The dHvA frequencies and effective masses for $\alpha$, $\beta$, and $\gamma$ sheets in GdRu$_{2}$Si$_{2}$ are close to those reported in LaRu$_{2}$Si$_{2}$ as shown in Table~\ref{tab}, \ref{tab2}.
The effective masses of LaRu$_{2}$Si$_{2}$ and GdRu$_{2}$Si$_{2}$ are in the order of 0.1-1$m_{0}$ and much smaller than that of the typical heavy-fermion system CeRu$_{2}$Si$_{2}$, in which heavy carriers with the effective mass of $> 100 m_{0}$ was observed \cite{doi:10.1143/JPSJ.61.3457}.
The light effective mass of GdRu$_{2}$Si$_{2}$ indicates that the $4f$ band is far below the Fermi level, which is consistent with the fact that the magnetization saturates at 7$\mu_{\mathrm{B}}$ per Gd${}^{3+}$ above $\SI{10}{\tesla}$ (see Fig.~\ref{Fig2}(a)).
The band calculation indeed predicts the 4$f$ band is located $\sim \SI{4}{\electronvolt}$ below the Fermi energy \cite{PhysRevLett.125.117204}.

\begin{table}
  \caption{Observed dHvA frequencies and corresponding cyclotron masses in GdRu$_{2}$Si$_{2}$ and LaRu$_{2}$Si$_{2}$ ($B \varParallel c$).}
\label{tab}
\begin{ruledtabular}
\begin{tabular}{ccccr}
 \multirow{2}{*}{Band}\rule[0mm]{5mm}{0mm} & \multicolumn{2}{c}{LaRu$_{2}$Si$_{2}$\cite{doi:10.1143/JPSJ.61.960}}\rule[0mm]{5mm}{0mm} & \multicolumn{2}{c}{GdRu$_{2}$Si$_{2}$} \rule[0mm]{1mm}{0mm} \\
  & $F$ (kT) & $m^{*}/m_{0}$\rule[0mm]{5mm}{0mm} & $F$ (kT) &\multicolumn{1}{l}{$m^{*}/m_{0}$} \rule[0mm]{7mm}{0mm}\\ \hline
$\alpha$\rule[0mm]{5mm}{0mm}  &  27.2 & 2.37\rule[0mm]{5mm}{0mm} & - & -\rule[0mm]{12mm}{0mm}  \\ \hline
\multirow{2}{*}{$\beta$}\rule[0mm]{5mm}{0mm}  & 0.86 & 0.55\rule[0mm]{5mm}{0mm} & 0.56 &  0.30\rule[0mm]{3mm}{0mm}($\beta_{\uparrow}$)\rule[0mm]{1mm}{0mm}\\
   & 1.35  & 0.67\rule[0mm]{5mm}{0mm}  & 0.76 &-\rule[0mm]{4mm}{0mm}  ($\beta_{\downarrow}$)\rule[0mm]{1mm}{0mm}  \\ \hline
\multirow{2}{*}{$\gamma$}\rule[0mm]{5mm}{0mm}    & \multirow{2}{*}{0.65} & \multirow{2}{*}{0.51}\rule[0mm]{5mm}{0mm} & 0.76 & 0.55\rule[0mm]{3mm}{0mm}($\gamma_{\downarrow}$)\rule[0mm]{1mm}{0mm} \\
  &  & & 1.16 & 0.58\rule[0mm]{3mm}{0mm}($\gamma_{\uparrow}$)\rule[0mm]{1mm}{0mm}\\
\end{tabular}
\end{ruledtabular}
\end{table}
\begin{table}
\caption{Observed dHvA frequencies and corresponding cyclotron masses in GdRu$_{2}$Si$_{2}$ and LaRu$_{2}$Si$_{2}$ ($B \varParallel a$).}
\label{tab2}
\begin{ruledtabular}
\begin{tabular}{ccccr}
\multirow{2}{*}{Band}\rule[0mm]{5mm}{0mm} & \multicolumn{2}{c}{LaRu$_{2}$Si$_{2}$\cite{doi:10.1143/JPSJ.61.960}}\rule[0mm]{5mm}{0mm} & \multicolumn{2}{c}{GdRu$_{2}$Si$_{2}$} \rule[0mm]{1mm}{0mm} \\
  & $F$ (kT) & $m^{*}/m_{0}$\rule[0mm]{5mm}{0mm} &  $F$ (kT) &\multicolumn{1}{l}{$m^{*}/m_{0}$} \rule[0mm]{7mm}{0mm}\\ \hline
 \multirow{2}{*}{$\alpha$\rule[0mm]{5mm}{0mm}} & \multirow{2}{*}{13.0} & \multirow{2}{*}{1.44\rule[0mm]{5mm}{0mm}} & 13.8 & -\rule[0mm]{4mm}{0mm}($\alpha_{\uparrow}$)\rule[0mm]{1mm}{0mm} \\
  & & & 15.3 & -\rule[0mm]{4mm}{0mm}($\alpha_{\downarrow}$)\rule[0mm]{1mm}{0mm} \\\hline
 \multirow{2}{*}{$\beta$}\rule[0mm]{5mm}{0mm}   &  1.43  & 0.53\rule[0mm]{5mm}{0mm} & 1.14 &-\rule[0mm]{4mm}{0mm}($\beta_{\uparrow}$)\rule[0mm]{1mm}{0mm}\\
    & 2.12  & 0.84\rule[0mm]{5mm}{0mm} & 1.29 & -\rule[0mm]{4mm}{0mm}($\beta_{\downarrow}$)\rule[0mm]{1mm}{0mm}\\ \hline
 $\gamma$\rule[0mm]{5mm}{0mm} & 0.75 & 0.47\rule[0mm]{5mm}{0mm} & 0.46 & -\rule[0mm]{4mm}{0mm}($\gamma_{\uparrow}$)\rule[0mm]{1mm}{0mm} \\
\end{tabular}
\end{ruledtabular}
\end{table}


\subsection{C. Magnetic interactions in GdRu$_{2}$Si$_{2}$}
Several theoretical frameworks have succeeded in reproducing the topological spin textures in centrosymmetric metal systems taking different magnetic interactions into account \cite{PhysRevB.103.104408, PhysRevB.103.064414, PhysRevLett.124.207201, Yambe2021, PhysRevB.104.184432, PhysRevLett.101.156402, PhysRevLett.108.096401, PhysRevLett.118.147205, hayami2017effective, PhysRevB.103.024439, PhysRevB.103.054422, PhysRevLett.125.117204, PhysRevLett.128.157206}.
Here, based on the experimental results shown above, major magnetic interactions in this system are discussed.

First, we point out the significance  of the RKKY interaction, the itinerant-electrons-mediated interaction, in GdRu$_{2}$Si$_{2}$.
According to the previous STM observation, different patterns in the local density of states were observed in each phase, reflecting the underlying magnetic structures \cite{yasui2020imaging}.
This indicates that the electronic band structure can be modified due to the different spin textures via the exchange interaction between the local moments and itinerant electrons.
The modification of the band structure may change the quantum oscillation frequencies, which is consistent with our observation of the change in dHvA frequencies at the transition from Phase III to the FF phase (Fig.~\ref{Fig2}).
The modulation of the deeply localized Gd-4$f$ spins can change the modulation pattern of the itinerant electrons at the Fermi level via the RKKY interaction.

On the one hand, the first-principles calculation study suggests that Fermi-surface nesting in GdRu$_{2}$Si$_{2}$ enhances the RKKY interaction that stabilizes the helical modulation along the nesting vector \cite{PhysRevLett.128.157206}.
The study predicts a barrel-shaped nested Fermi surface located around the zone center, whose nesting vector matches the experimentally observed magnetic modulation $\mathbf{Q}=(0.22, 0, 0)$.
Although we do find a barrel-shaped sheet (pillar-like part of the $\gamma$ band), it resides at the zone corner.
The predicted nesting period, however, shows a good agreement with the radius of the band $\beta$.
Assuming a circular cross-section, the magnetic modulation $\mathbf{Q}=(0.22, 0, 0)$ corresponds to the cross-section whose radius is $k \sim \SI{0.17}{\angstrom ^{-1}}$.
This is close to the estimated Fermi wavevectors of the band $\beta$ in the FF phase (0.13 and $\SI{0.15}{\angstrom^{-1}}$, respectively (Table.~\ref{tab0})).
The fair agreement with the experimental value might indicate that the helimagnetic modulation can be enhanced by the Fermi-surface effect in GdRu$_{2}$Si$_{2}$.

We also comment on the interorbital frustration mechanism proposed by the other first-principles calculation study \cite{PhysRevLett.125.117204}.
This study starts with the band structure calculation assuming the ferromagnetic background and obtain $J(\mathbf{Q})$, which is then decomposed into each atomic component.
Although we cannot discuss the mechanism they propose from our experimental results, the agreement between the present experimental results and theoretical calculation strongly indicates the plausibility of the calculated electronic structures presented in Ref. \cite{PhysRevLett.125.117204}.

\section{Conclusion}
In this research, we performed magnetic torque and magnetoresistance measurements on the centrosymmetric skyrmion host GdRu$_{2}$Si$_{2}$.
We observed dHvA oscillations in both Phase III and the FF phase, and SdH oscillations in the FF phase.
The angular dependence of the dHvA in the FF phase is qualitatively reproduced by the band structure calculation.
This indicates that the previous $ab$ $initio$ calculation \cite{PhysRevLett.125.117204} well captures the electronic structure and magnetic interaction in this system.

Fermi surface reconstruction was also observed as a change in the dHvA-oscillation frequency, reflecting the coupling between conduction electrons and localized spins.
This confirms the significance of the RKKY interaction in GdRu$_{2}$Si$_{2}$.
The Fermi wavevectors corresponding to the $\beta$ surface detected by the dHvA oscillations in the FF phase are almost the same as that of the wavevector of magnetic modulation, as was predicted by another microscopic calculation \cite{PhysRevLett.128.157206}.
This agreement might suggest the Fermi surface nesting, stabilizing the magnetic modulation characterized by the nesting vector with the RKKY interaction.

As we mentioned in the previous section, the Fermi surface reconstruction might occur at every magnetic phase transition.
Thus, the electronic structures in the SkL phase (Phase II) might deviate from those of Phase III and the FF phase.
Indeed, the real-space observation of conduction electrons by the STM experiment revealed the different patterns in the local density of states corresponding to each magnetic phase from Phase I to III and the FF phase \cite{yasui2020imaging}.
As a probe to investigate the electronic structure, angle-resolved photoemission spectroscopy (ARPES) is one of the powerful techniques, which was recently conducted and unveiled the electronic structure of Phase I \cite{ishizakalab}.
However, the measurement can usually be conducted only at zero field, and the electronic structure of Phase II is left for future research.

\section{Acknowledgement}
The author thanks T.-H. Arima, S. Hayami, K. Ishizaka, and Y.Yasui for the insigthfull discussions.
This work was partly supported by JSPS KAKENHI, Grants-In-Aid for Scientific Research (Nos. 19H05825, 20K14403, 20K20892, 21H04437, 21H04990, 21H05470, 21K13873, 21K13876, and 22H00104), and JST PRESTO (Nos. JPMJPR20B4 and JPMJPR20L7).
N.M. thanks the World-leading Innovative Graduate Study Program for Materials Research, Information, and Technology (MERIT-WINGS).
%
\bibliography{bib}

\end{document}